\documentstyle[prb,aps]{revtex}
\begin{document}
\twocolumn[\hsize\textwidth\columnwidth\hsize\csname
@twocolumnfalse\endcsname
\draft
\title{Crystallization of a classical two-dimensional electron system:\\
  Positional and orientational orders}
\author{Satoru Muto and Hideo Aoki}
\address{Department of Physics, University of Tokyo, Hongo,
  Tokyo 113-0033, Japan}
%\date{\today}
\date{Physical Review B{\bf 59}, 14911 (1999)}
\maketitle
\begin{abstract}
Crystallization of a classical two-dimensional one-component plasma
(electrons interacting with the Coulomb repulsion in a uniform
neutralizing positive background) is investigated with a
molecular-dynamics simulation.  The positional and the orientational
correlation functions are calculated, to the best of our knowledge,
for the first time.  We have found an indication that the solid phase
has a quasi-long-range (power-law) positional order along with a
long-range orientational order.  This indicates that, although the
long-range Coulomb interaction is outside the scope of Mermin's
theorem, the absence of ordinary crystalline order at finite
temperatures applies to the electron system as well.  The `hexatic'
phase, which is predicted between the liquid and the solid phases by
the Kosterlitz-Thouless-Halperin-Nelson-Young theory, is also
discussed.
\end{abstract}
%\pacs{73.20.Dx}
\vskip2pc]

Wigner pointed out in the 1930's that an electron system should
crystallize due to the Coulomb repulsion for low enough
densities.~\cite{wigner} Although quantum effects are essential, the
concept of electron crystallization can be generalized to classical
cases.  In fact, Grimes and Adams succeeded in observing a
liquid-to-solid transition in a classical two-dimensional (2D)
electron system on a liquid-helium surface in
1979.~\cite{grimes-adams} In this system electrons obey classical
statistics because the Fermi energy is much smaller than $k_{B}T$.
The thermodynamic properties of the classical electron system are
wholly determined by the dimensionless coupling constant $\Gamma$, the
ratio of the Coulomb energy and the kinetic energy.  Here $\Gamma
\equiv (e^{2}/4\pi\epsilon a)/k_{B}T$, where $a=(\pi n)^{-1/2}$ is the
mean-distance between the electrons, $n$ the density of electrons, $e$
the charge of an electron, and $\epsilon$ is the dielectric constant
of the substrate.  For $\Gamma \ll 1$ the system will behave as a gas
while for $\Gamma \gg 1$ as a solid.  Grimes and
Adams~\cite{grimes-adams} found that the phase transition occurs at
$\Gamma_{c} = 137 \pm 15$.

On the other hand, Mermin proved rigorously, more than thirty years
ago, that no true long-range crystalline orders are possible in the
thermodynamic limit at finite temperatures in 2D.~\cite{mermin} In the
proof, short-range interactions are assumed, while the $1/r$ Coulomb
interaction is too long-ranged to apply Mermin's arguments (see
Ref.~\onlinecite{mermin} for more precise mathematical conditions).
Although there have been some theoretical
attempts~\cite{chakravarty,alastuey} to extend the theorem to the
long-range Coulomb case, no rigorous proof is attained up to now.
Numerically, Gann et~al.~\cite{gann} investigated the problem with a
Monte Carlo method by calculating the root-mean-square displacement.
However they found it difficult to rule out the possibility that the
root-mean-square displacement approaches a constant value in the
thermodynamic limit, which is a conventional definition of a solid.

Another intriguing problem concerning 2D systems is the melting
mechanism, where the Kosterlitz-Thouless-Halperin-Nelson-Young (KTHNY)
theory~\cite{halperin-nelson,young} has predicted the existence of the
`hexatic' phase between the liquid and the solid phases (see
Refs.~\onlinecite{halperin,nelson,strandburg} for reviews).  The
hexatic phase is characterized by a short-range positional order and a
quasi-long-range orientational order.  Numerical calculations have
been carried out by several authors with a molecular-dynamics (MD)
method.~\cite{morf,kalia,hockney-brown} Most of the authors have
obtained the melting point in good agreement with the Grimes-Adams
experiment.  However the results are rather controversial on the
verification of the hexatic phase.  Morf~\cite{morf} concluded, from a
MD result for the shear modulus, that the result agrees well with the
KTHNY theory.  On the other hand, Kalia et~al.~\cite{kalia} observed
hysteresis in the temperature dependence of the total energy in their
MD simulation to conclude that the melting is a first-order phase
transition, which is incompatible with the KTHNY theory.  The Monte
Carlo study by Gann et~al.~\cite{gann} is also indicative of a
first-order phase transition.

In addressing both of the above problems, i.e., the range of the
ordering in the solid phase and the nature of the melting, the most
direct way is to investigate the positional and the orientational {\it
correlation functions}, since the phases should be characterized in
terms of them.  This is exactly the purpose of the present paper,
which is done, to the best of our knowledge, for this system for the
first time.

In order to accurately incorporate the temperature, we have employed
Nos\'{e}-Hoover's {\it canonical} MD method~\cite{nose,hoover} for the
classical 2D electron system, while the previous calculations were
done for micro-canonical ensemble.~\cite{morf,kalia,hockney-brown}
Electrons are confined to a rectangle with a rigid uniform
neutralizing positive background.  We impose periodic boundary
conditions.  An electron then interacts with infinite arrays of the
periodic images with the long-range $1/r$ potential.  We employ the
Ewald summation method to take care of this.  The rectangle is chosen
to be close to a square to minimize surface effects.  The aspect ratio
of the rectangle is taken to be $L_{y}/L_{x}=2/\sqrt{3}$, which can
accommodate a perfect triangular lattice~\cite{bonsall} with
$N=4M^{2}$ ($M$: an integer) particles.  The equations of motion are
integrated numerically with Gear's predictor-corrector algorithm.  We
adopt a time step of $1.0\times 10^{-12}$ sec, which guarantees
six-digit accuracy in the energy conservation after several tens of
thousands of steps.

We have performed the simulation both from a typical liquid ($\Gamma =
60$) and from a typical solid ($\Gamma = 200$).  The initial
conditions are set as follows: The electrons are placed randomly in
the liquid phase or placed at the perfect triangular lattice points in
the solid phase.  The velocities of the electrons are assigned
according to the Maxwell-Boltzmann distribution in either case.  The
lowest energy configurations are sought with a simulated annealing
method.  Namely, the positions and the velocities of the electrons are
updated for a certain time interval.  Once a thermal equilibrium sets
in, the temperature is raised or lowered by a small amount.  The
latest positions and velocities are used as the initial conditions for
the simulation at the new temperature.  This procedure temporarily
puts the system out of equilibrium, but updating the positions and
velocities for a certain time interval equilibrates the system.  This
annealing process is repeated from a liquid phase to a solid phase (or
vice versa) across the transition.  We take care that the system is
well equilibrated, especially near and after the transition, by
allowing large numbers of time steps.  The results presented in this
paper are for $N=900$ electrons with MD runs from 30,000 to 110,000
time steps for each value of $\Gamma$.  The correlation functions are
calculated for the last 20,000 time steps.

Following Cha and Fertig,~\cite{cha-fertig} we define the positional
and the orientational correlation functions from which we identify the
order in each phase.  First, the positional correlation function is
defined by
\begin{eqnarray}
  C(r) & \equiv & \langle \rho^{*}_{\bf G}({\bf r})
  \rho_{\bf G}({\bf 0}) \rangle
  \label{pos-corr-def-1} \\
  & = & \left\langle \frac{{\displaystyle \sum_{i,j}}
    \delta (r - |{\bf r}_{i} - {\bf r}_{j}|)
    \frac{1}{6}{\displaystyle \sum_{{\bf G}}}
    e^{{\rm i}{\bf G} \cdot ({\bf r}_{i} - {\bf r}_{j})}}
  {{\displaystyle \sum_{i,j}}
    \delta (r - |{\bf r}_{i} - {\bf r}_{j}|)} \right\rangle ,
  \label{pos-corr-def-2}
\end{eqnarray}
where ${\bf G}$ is the reciprocal vector of the triangular lattice,
and $\rho_{\bf G}({\bf r}) = \exp({\rm i} {\bf G} \cdot {\bf r})$.
The angular brackets in Eq.~(\ref{pos-corr-def-1}) stand for both the
summation over particles and the thermal average.  In
Eq.~(\ref{pos-corr-def-2}) a summation is taken over six reciprocal
vector ${\bf G}$'s that give the first peaks of the structure factor.
In practice, the $\delta$-function must be broadened so that it can be
handled numerically.

The orientational correlation function is defined by
\begin{eqnarray}
  C_{6}(r) & \equiv & \langle \psi^{*}_{6}({\bf r})
  \psi_{6}({\bf 0}) \rangle
  \label{ori-corr-def-1} \\
  & = & \left\langle \frac{{\displaystyle \sum_{i,j}}
    \delta (r - |{\bf r}_{i} - {\bf r}_{j}|)
    \psi^{*}_{6} ({\bf r}_{i}) \psi_{6} ({\bf r}_{j})}
  {{\displaystyle \sum_{i,j}} \delta (r - |{\bf r}_{i} - {\bf r}_{j}|)}
  \right\rangle,
  \label{ori-corr-def-2}
\end{eqnarray}
where $\psi_{6} ({\bf r}) = \frac{1}{n_{c}}\sum_{\alpha}^{\rm n.n.}
e^{6{\rm i}\theta_{\alpha}({\bf r})}$, and $\theta_{\alpha}({\bf r})$
is the angle of the vector connecting an electron at ${\bf r}$ and the
$\alpha$-th nearest neighbor with respect to a fixed axis, say, the
$x$-axis.  The summation is taken over $n_{c}$ nearest neighbors which
are determined by the Voronoi diagram~\cite{allen}, or equivalently
its dual mapping, the Delaunay triangulation.

We first look at the positional and the orientational correlation
functions for $\Gamma = 200$ and $\Gamma = 160$, typical solid phases,
in Fig.~\ref{corr-solid}.  Although $\Gamma \: (T)$ is high (low)
enough, the positional correlation is seen to decay slowly in both
cases, indicating an algebraic decay at large distances.  The
round-off in the correlation function around half of the linear
dimension of the system size is considered to be an effect of the
periodic boundary conditions.  The algebraic decay of the positional
correlation function implies that the 2D electron solid has only a
{\it quasi-long-range} positional order at finite
temperatures.~\cite{footnote1} Thus we have obtained a numerical
indication that Mermin's theorem~\cite{mermin} applies to the electron
system as well, which is consistent with the analytical (but not
rigorous) results obtained in Refs.~\onlinecite{chakravarty,alastuey}.

By contrast, the orientational order is seen to be long-ranged.
Therefore while the 2D electron solid has no true long-range
crystalline order, it is {\it topologically} ordered.  The triangular
structure is seen as the peaks in both the positional and the
orientational correlation functions (see the inset of
Fig.~\ref{corr-solid}).

We have plotted the Delaunay triangulation of a snapshot of the
electron configuration for a solid ($\Gamma = 160$) or for a liquid
($\Gamma = 90$) in Fig.~\ref{defect}.  We can in particular look at
the topological defects, i.e., five-fold and seven-fold coordinated
electrons.  From the result the defects are seen to appear in isolated
pairs, or more precisely in quartets, in the solid phase, which
explains how a quasi-long-range positional order is compatible with a
long-range orientational one.  On the other hand, defects appear with
a high density in the liquid phase.

We now focus on the orientational correlation function near the
crystallization in Fig.~\ref{ori-corr-near-cryst}.  The result is
obtained by cooling the system from a liquid to a solid.  The
orientational order is short-ranged for $\Gamma \leq 120$ and
long-ranged for $\Gamma \geq 140$.  Around the liquid-solid boundary
($\Gamma = 130$), the orientational correlation function, plotted on a
double logarithmic scale in Fig.~\ref{ori-corr-near-cryst}, indicates
an algebraic decay (while the positional order is short-ranged). The
power evaluated from the data is approximately equal to unity, which
is greater than the upper bound of $1/4$ predicted by the KTHNY
theory.  Large statistical errors near the transition, however,
prevent us from drawing any definite conclusion on the existence of
the hexatic phase.  In fact, the correlation functions behave like a
solid at $\Gamma = 130$ when the system is heated from a solid to a
liquid with no indication of the hexatic phase.  This may be due to a
finite-size effect, where a solid can be pinned in a melting process.
For a small system, $N=100$, a solid phase in fact persists down to
$\Gamma = 120$ when heated.

Another finite-size effect is that, when the system crystallizes, the
crystal axes can tilt from the unit cell axes of the finite system.
The crystallization does occur in a tilted way in the present
simulation.  The misalignment causes a long relaxation time for the
system to reach the lowest energy state.  However the fact that the
system crystallizes with tilted axes shows in itself that the $N=900$
system is sufficiently large in that boundary effects are not too
strong.  By contrast, we found that the crystal axes are always
aligned to the unit cell for $N=100$.  For the $N=100$ system, which
is the size employed by Kalia et~al.~\cite{kalia}, we found no
indication of the hexatic phase, either.  A numerical difficulty in MD
simulations also arises from finite time steps.  Finite time effects
might result in insufficient equilibration, especially near a
continuous transition.  Even if the system is well-equilibrated, it
would be difficult to tell a slow exponential decay from an algebraic
decay in the correlation function for a finite system.

In summary, we have performed a molecular-dynamics simulation to
investigate the ordering of a classical 2D electron system.  From the
positional and the orientational correlation functions we have found
an indication that there is a quasi-long-range positional order and a
long-range orientational order in the solid phase, which implies that
Mermin's theorem is not spoiled even for the long-range $1/r$
interaction.  On the other hand, we have obtained only a sign, not a
conclusive result, for the existence of the hexatic phase predicted by
KTHNY theory, which thus remains an open question.

Although we have basically electrons on a liquid-helium surface in
mind, a planar classical one-component plasma is recently realized as
laser-cooled ions trapped in a disk region.~\cite{mitchell} The disk
has a finite thickness, for which stable crystalline phases are
observed.  If the ions could be trapped completely in 2D, the present
picture would be applicable.  Conversely, it is an interesting
theoretical problem to extend the present line of approach to planar
systems with finite thicknesses.

We wish to thank Kazuhiko Kuroki, Hiroshi Imamura, Katsunori Tagami,
and Naruo Sasaki for valuable discussions.
The numerical calculations were mainly done with Fujitsu VPP500 at
the Supercomputer Center, Institute for Solid State Physics,
University of Tokyo.

%\newpage
\begin{figure}
  %\epsfile{file=Fig1.eps,scale=0.7}
  %\epsfile{file=Fig1.eps,scale=0.45}
  \caption{The positional (the upper panel) and the orientational
    (the lower panel) correlation functions for $\Gamma = 200$
    and $\Gamma = 160$.
    The horizontal scale is in units of the lattice
    constant of the triangular lattice.
    The inset in the lower panel illustrates the first four distances
    which give the peaks in the correlation functions.}
  \label{corr-solid}
\end{figure}

\begin{figure}
  %\epsfile{file=Fig2a.eps,scale=0.45}
  %\epsfile{file=Fig2a.eps,scale=0.4}

  %\epsfile{file=Fig2b.eps,scale=0.45}
  %\epsfile{file=Fig2b.eps,scale=0.4}
  \caption{The Delaunay triangulation of a snapshot of the electron
    configuration for a solid with $\Gamma =160$ (a)
    or for a liquid with $\Gamma = 90$ (b).
    Five(seven)-fold coordinated electrons are marked with
    empty (solid) circles.
    In (b) four(eight)-fold coordinated electrons are marked with
    empty (solid) squares.
    A part of the system is shown in either panel.}
  \label{defect}
\end{figure}

\begin{figure}
  %\epsfile{file=Fig3.eps,scale=0.7}
  %\epsfile{file=Fig3.eps,scale=0.5}
  \caption{The orientational correlation function near the crystallization.
    The result was obtained by cooling the system from a liquid
    to a solid.}
  \label{ori-corr-near-cryst}
\end{figure}


\begin{references}
\bibitem{footnote1} 
A power-law decay of the positional correlation function also implies 
that the Bragg spots in the structure factor should have 
a hexagonal symmetry with a power-law divergence of each spot 
in the thermodynamic limit.
\end{references}
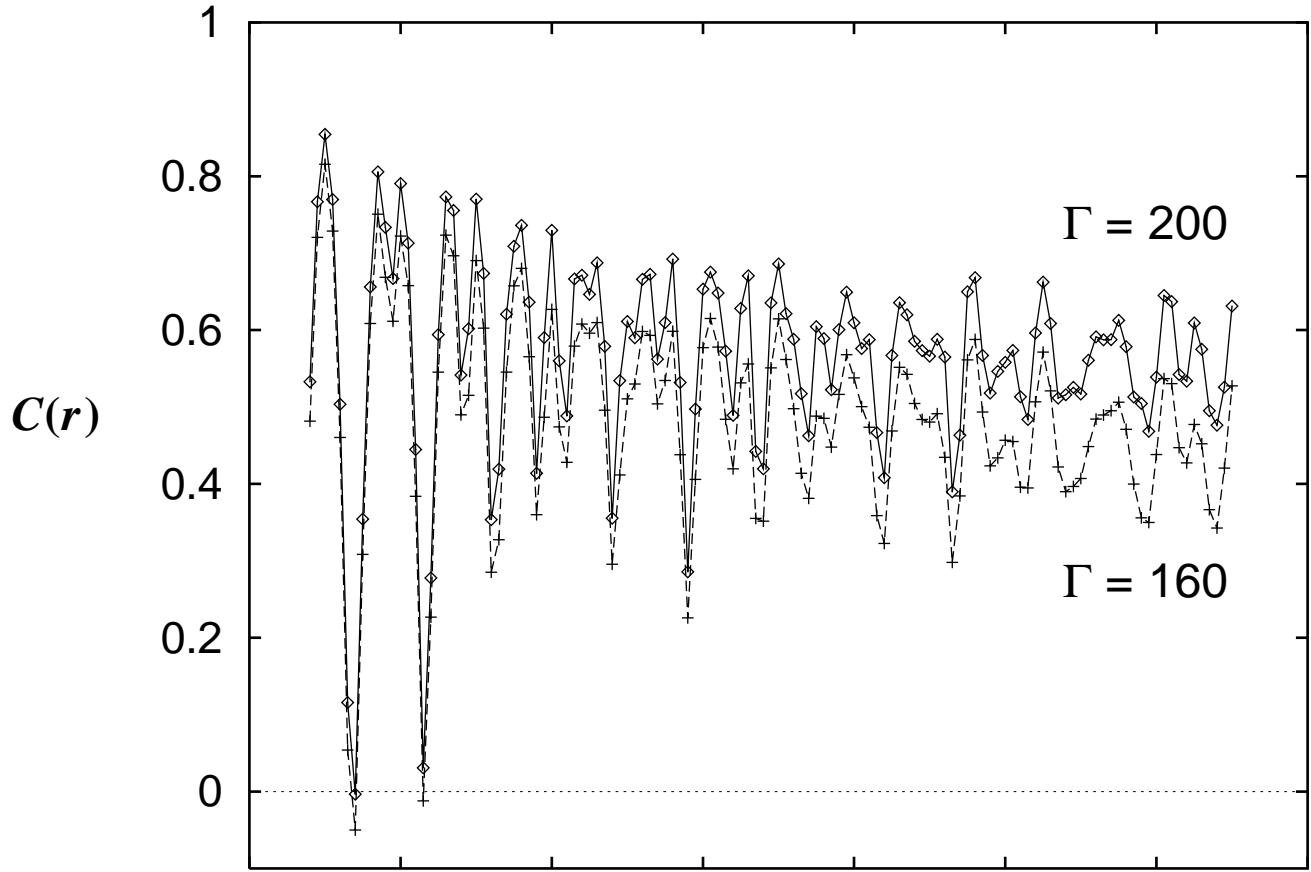
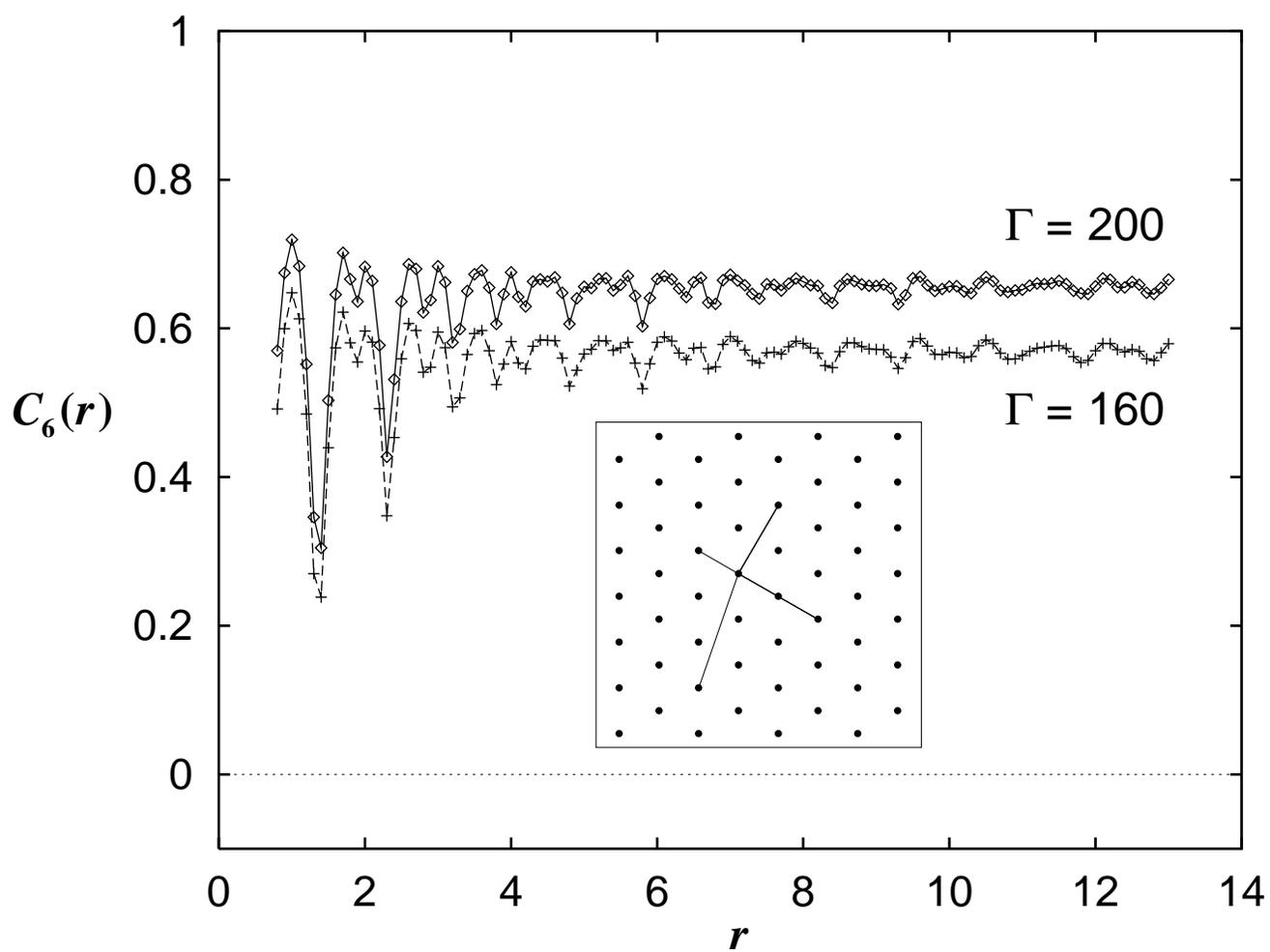
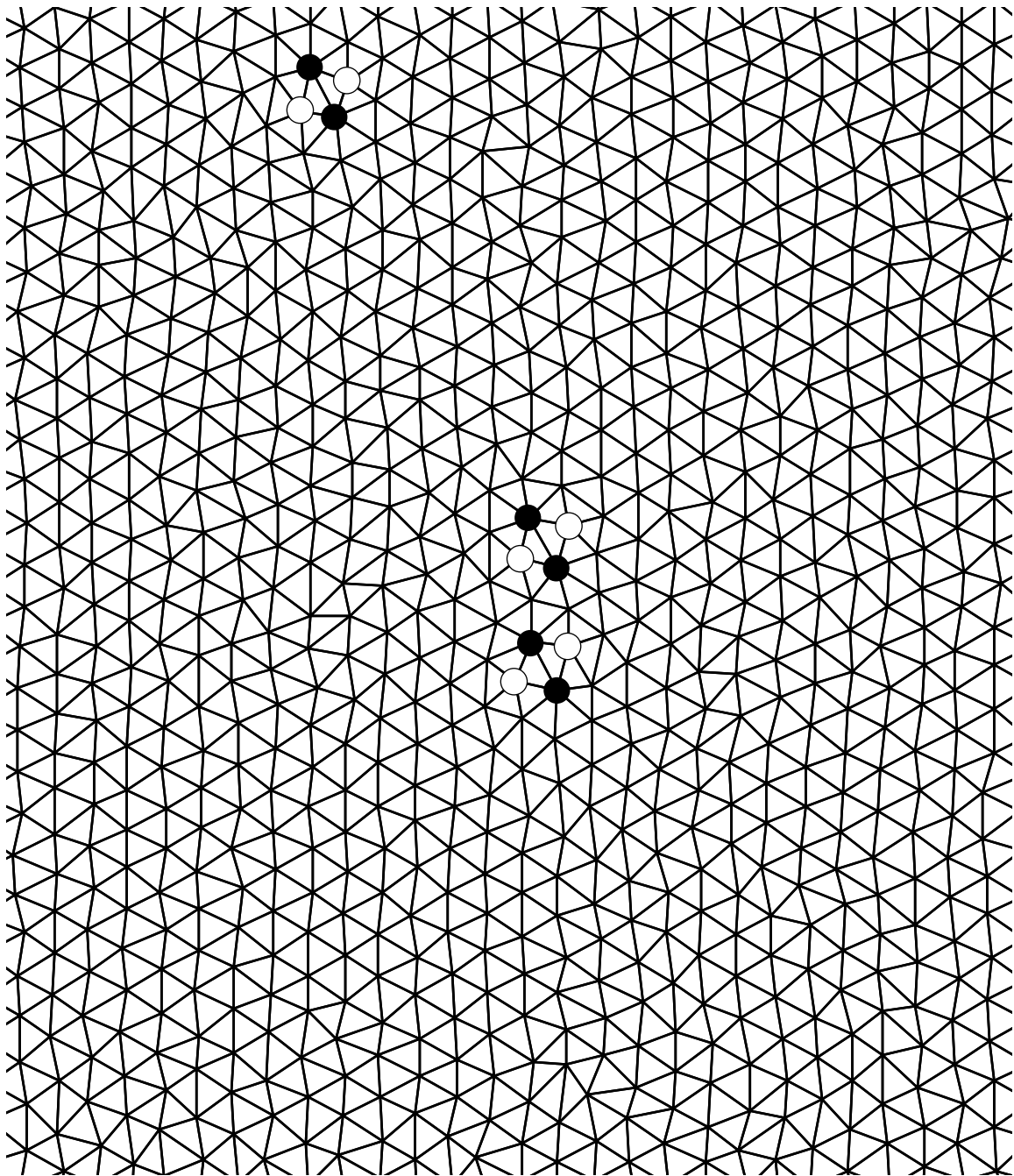
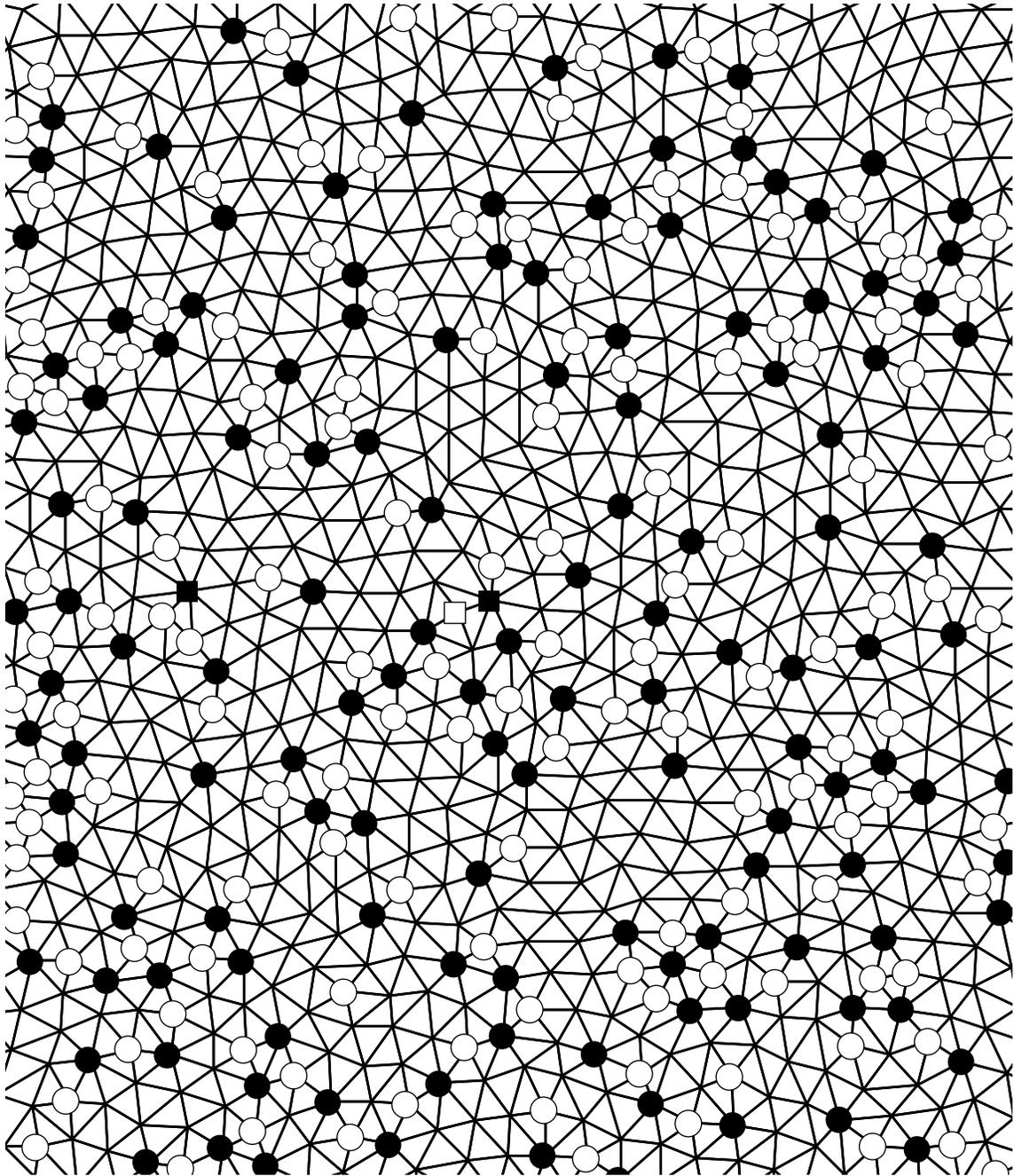
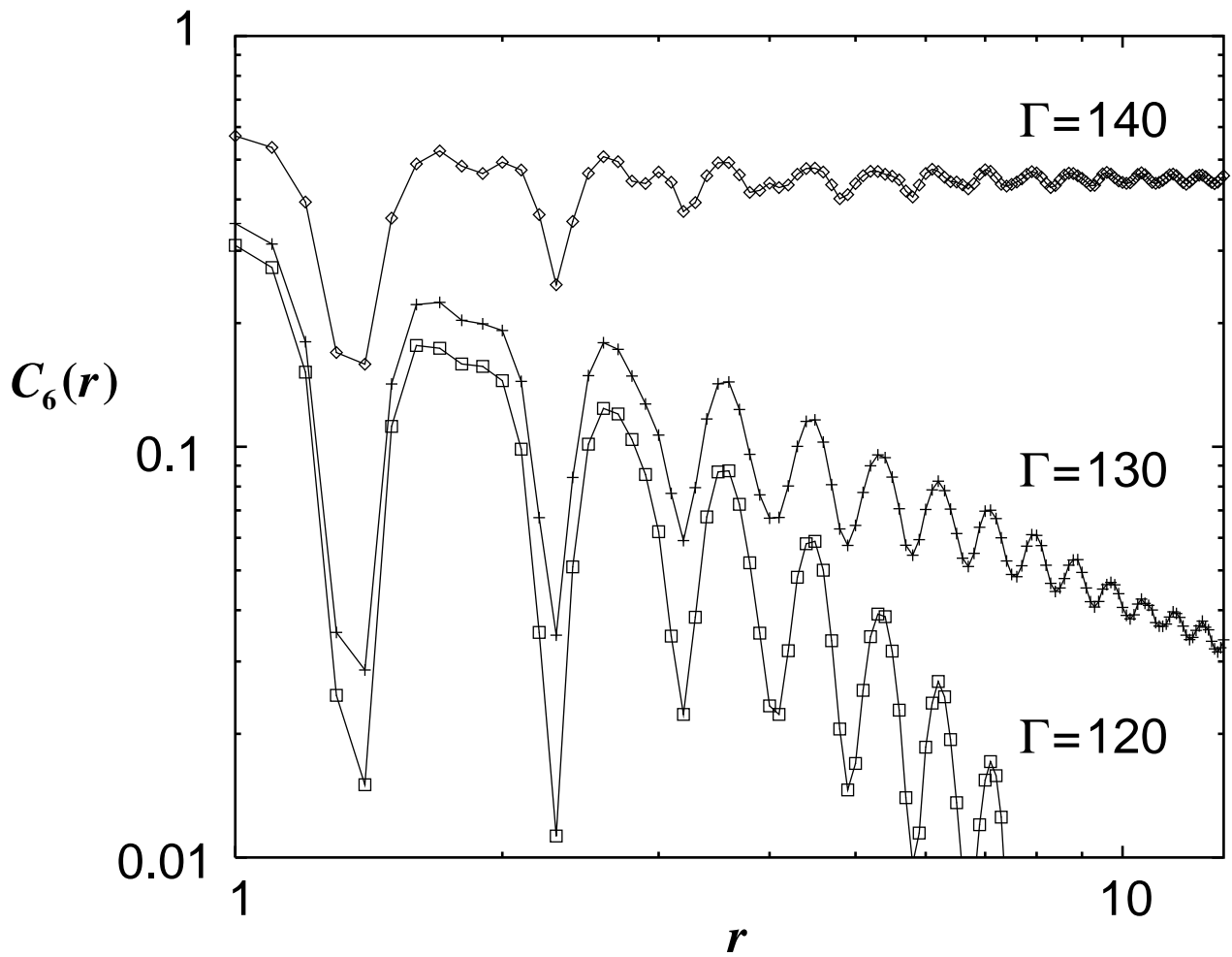
\end{document}